\documentclass[preprint,prc,showpacs,preprintnumbers,
                unsortedaddress,amsmath,amssymb,floatfix]{revtex4}
\usepackage{graphicx}
\usepackage{dcolumn}
\usepackage{bm}
\usepackage{longtable}
\usepackage{color}
\usepackage[bookmarksnumbered,plainpages]{hyperref}
 \begin{document}
 \title{Application of the RMF mass model to the $r$-process and the influence of mass
 uncertainties}
\author{B. Sun$^{1,2}$\footnote{B.Sun@gsi.de}, F. Montes$^{2,3}$, L.S. Geng$^{1,4}$, H. Geissel$^2$, Yu.A. Litvinov$^2$ and J.
Meng$^{1,5,6,7}$ \footnote{mengj@pku.edu.cn}}

\affiliation{ $^{1}$School of Physics and State Key Laboratory of Nuclear Physics and Technology,
Peking University, 100871 Beijing, China  \\
$^{2}$Gesellschaft f\"ur Schwerionenforschung GSI, 64291 Darmstadt,
Germany \\
$^{3}$National Superconducting Cyclotron
Laboratory, Michigan State University, E. Lansing, MI 48824, USA\\
$^{4}$Departamento de F\'{\i}sica Te\'orica and IFIC, Centro Mixto,
Institutos de Investigaci\'on de Paterna - Universidad de
Valencia-CSIC \\
$^{5}$ Department of Physics, University of Stellenbosch,
Stellenbosch, South Africa \\
$^{6}$Institute of Theoretical Physics, Chinese Academy of Sciences,
Beijing, China \\
$^{7}$Center of Theoretical Nuclear Physics,
National Laboratory of Heavy Ion Accelerator, 730000 Lanzhou, China}

 \date{\today}

\begin{abstract}
\baselineskip=13pt A new mass table calculated by the relativistic
mean field approach with the state-dependent BCS method for the
pairing correlation is applied for the first time to study
$r$-process nucleosynthesis. The solar $r$-process abundance is well
reproduced within a waiting-point approximation approach. Using an
exponential fitting procedure to find the required astrophysical
conditions, the influence of mass uncertainty is investigated.
$R$-process calculations using the FRDM, ETFSI-Q and HFB-13 mass
tables have been used for that purpose. It is found that the nuclear
physical uncertainty can significantly influence the deduced
astrophysical conditions for the $r$-process site. In addition, the
influence of the shell closure and shape transition have been
examined in detail in the $r$-process simulations.
\end{abstract}

 \maketitle
\section{Introduction}

  It is of the utmost interest to explore the ``terra incognita'' of exotic nuclei,
  as evidenced by the fact that several Radioactive Ion Beam (RIB) facilities are being
  upgraded, under construction or planned to be constructed worldwide.
  Such investigations of the properties of these
  exotic nuclei, which may behave very differently from the nuclei around the $\beta$-stability line,
  result in new discoveries such as the halo phenomenon~\cite{THH-PRL85,Schwab94} - nucleons spread
  like a thin mist around
  the nucleus, which can significantly increase the nuclear reaction ratio. Stellar
  nucleosynthesis processes such as the $r$-process~\cite{Cowan-PR91,Qian-PPNP03},
  which is responsible for roughly half of the enrichment of elements heavier than
  iron in the universe, also require a thorough understanding of the properties
  of exotic nuclei. Key properties like masses for example, determine the path that
  the  nucleosynthesis process follow in the nuclei chart.
  Nevertheless, despite many experimental efforts, present knowledge of exotic nuclei still does not include much
  of what is required for a complete understanding of $r$-process nucleosynthesis.
  After the first systematic introduction to the $r$-process~\cite{BBFH} half a century ago,
  $r$-process calculations for a long time could only rely on the phenomenological
  nuclear droplet mass formula~\cite{Hilf} because of the lag of both experimental
  and theoretical development. Fortunately, in the last 15 year the theoretical
  study of nuclear properties has made tremendous progress and $r$-process
  calculations~\cite{Kartz-APJ93,pfeiffer-ZPA97,Wanajo-APJ04} have been
  carried out based on the refined droplet model FRDM~\cite{FRDM},
  Hartree-Fock approach like ETFSI-Q~\cite{ETFSIQ}, and the very recent microscopic
  rooted Hartree-Fock Bogliubov (HFB)~\cite{HFB,HFB2,Samyn-PRC03}.

  Despite the progress in the theoretical nuclear structure physics, mass models
  predictions (which by design concentrate on different nuclear structure aspects)
  still show a large deviation when going to very neutron-rich nuclides, even though
  they have achieved similar quality to describe known nuclides. This is specially
  troublesome since the astrophysical scenario in which an $r$-process may
  occur is a matter of debate and all astrophysical simulations dedicated to
  the nature of the stellar environment depend on the input from nuclear physics.
  Mass model predictions, even in models that give similar global $rms$ error still show
  local deviations differently.

  In principle, microscopic-rooted mass models should have a more
  reliable extrapolation to the unknown regions, therefore
  these studies have received more and more interest as evidenced by the
  increasing number of non-relativistic HFB investigations~\cite{HFB,HFB2,Samyn-PRC03,HFB-13}.
  Based on a mass-driven fitted method, the latest HFB models have achieved a similar
  quality ($rms\sim$ 0.7 MeV) as the phenomenological FRDM mass model for known masses.
  More recently, another microscopic-rooted approach, the relativistic
  mean-field (RMF) theory~\cite{Wale-AP74} has
  received broad attention due to its successful description of several
  nuclear phenomena during the past years (for recent reviews,
  refer to Refs.~\cite{Bender-RMP03,Meng06}). In the framework of
  the RMF theory, the nucleons interact via the exchanges of mesons
  and photons. The corresponding large scalar and vector fields, of
  the order of a few hundred MeV, provide simple and efficient
  descriptions of several important phenomena such as the large
  spin-orbit splitting, the density dependence of optical potential,
  the observation of approximate pseudo-spin symmetry, etc..
  Moreover, the RMF theory can reproduce well the isotopic shifts
  in the Pb region~\cite{SLR-PLB93}, explain naturally the origin
  of the pseudo-spin symmetry~\cite{AHS-PLB69,HA-69} as a
  relativistic symmetry~\cite{Ginocchio-97,MSYP-PRC98,MSYA-PRC99,Chen-CPL03} and
  spin symmetry in the anti-nucleon spectrum~\cite{ZMP-PRL03}.

  The first RMF mass table was reported in Ref.~\cite{Hirata-NPA97} for 2174 even-even nuclei
  with $8\leq Z \leq 120$ but without including pairing correlations.
  Later on, the calculation was improved by adopting a constant-gap BCS
  method and calculated 1200 even-even nuclei
  with $10 \leq Z \leq 98$~\cite{Lalazissis-ADNDT99}, most of which are
  close to the $\beta$-stability line. More recently,
  using the state-dependent BCS method with a $\delta$-force~\cite{Sanulescu-PRC00,Geng-PTP03},
  the first systematic study of the ground state
  properties of over 7000 nuclei ranging from the proton drip line to the neutron drip line was
  performed \cite{RMFBCS}. Comparison of this calculation with experimental data and to the
  predictions of other mass models will be presented in more detail in Sec II.

  Considering the recent development of the microscopic mass models in both the HFB and RMF approach,
  it is very interesting to examine their applicability to an
  $r$-process calculation. The main goals of this paper is to explore to what extent the
  solar $r$-process abundance can be reproduced using the new RMF
  mass table~\cite{RMFBCS} and by comparing with
  other theoretical mass models, to determine the influence of nuclear mass
  uncertainty in $r$-process calculations. The paper is organized as follows. In
  Sec.~\ref{sec:RMF} the global agreement of the new RMF mass table
  with the experimental data is discussed and the RMF
  prediction in the very neutron-rich range is compared with the
  FRDM~\cite{FRDM}, the ETFSI-Q~\cite{ETFSIQ} and the latest
  HFB-13~\cite{HFB-13} mass tables. In Sec.~\ref{sec:$r$-process},
  a short introduction to a site-independent $r$-process approach is
  given. In Sec.~\ref{sec:calculation}, the new mass table is
  applied to reproduce the solar $r$-process abundances. In
  addition, the result is compared to the $r$-process
  abundances obtained with the predictions of the FRDM, ETFSI-Q and
  HFB-13 mass models. Finally the summary and conclusions are given
  in Sec.~\ref{sec:conclusion}.

  \section{Global behavior of the new RMF mass table}
    \label{sec:RMF}
\begin{figure}
  \includegraphics[scale=1]{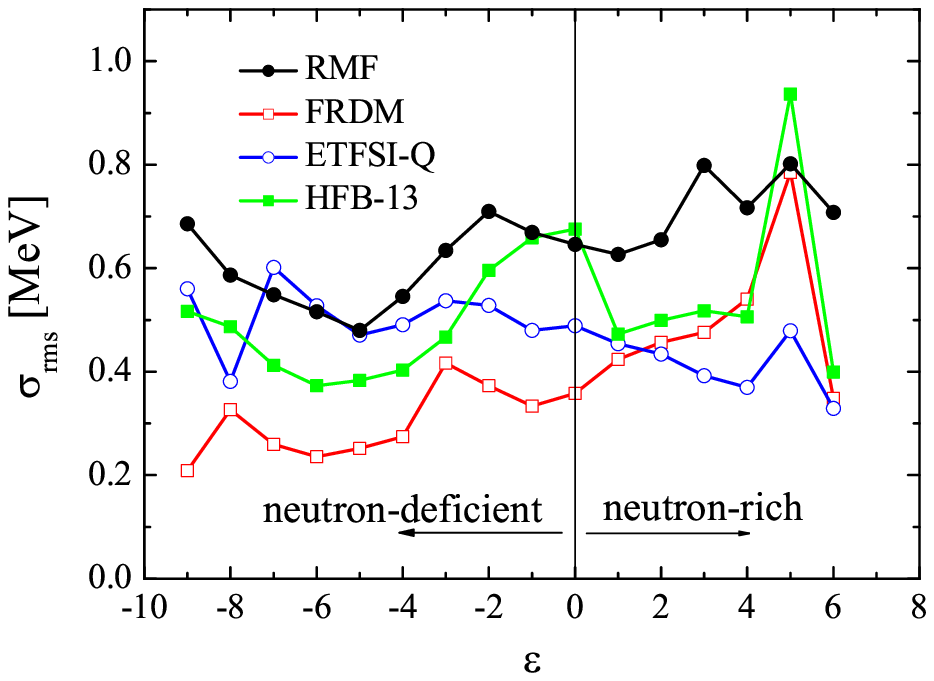}
     \caption{(Color online) The $rms$ deviation $\sigma_{rms}$ of one-neutron separation energy $S_n$
     with respect to experimental data~\cite{AW03}
     as a function of the distant from the $\beta$-stability line $\varepsilon=Z_0-Z$ for different mass models,
     where $Z_0$ stands for the proton number of the
     most stable isotope in the isobaric chain with mass number $A$.}
     \label{fig:sn_sigma}
  \end{figure}

  With about 10 parameters fitted to the ground-state properties of around 10 spherical nuclei,
  the RMF approach with the TMA parameter set is found to give a satisfactory description for all
  the nuclei in the nuclear chart.
  The model deviation of one-neutron separation energy $S_n$ with respect to the known experimental
  data can be characterized by the $rms$ deviation ($\sigma_{rms}$)
  \begin{equation}
   \sigma_{rms}=\sqrt{\frac{1}{n}\sum_{i=1}^n (S_{n}^{\mathrm{th}}-S_{n}^{\mathrm{exp}})_i^2}  \;,
   \label{eq:rms}
  \end{equation}
  Although a relatively large $rms$ deviation for the absolute mass value is found for the RMF
  calculation in comparison with the FRDM and HFB-13 models, the finite differences in binding
  energies such as the practical used one-neutron separation energies $S_n$ are well
  predicted due to the cancelation of systematic error \cite{RMFBCS, Geissel06}. The $rms$
  deviation of $S_n$ for the FRDM,
  ETFSI-Q, HFB-13 and RMF models with respect to experimentally values~\cite{AW03} are 399
  keV, 528 keV, 546 keV and 654 keV, respectively. Here the comparisons include nuclei with $Z,
  N\geq 8$. Comparing the predictions of the RMF model to the known
  values~\cite{AW03} results in discrepancies between -1.4 MeV to 1
  MeV, while the difference between either the FRDM or the HFB-13
  and the experimental data is in the range of -1.3 MeV to 0.5 MeV.
  It shows that the microscopic model such as the RMF approach can
  almost achieve the same level of accuracy for known one-neutron separation energy $S_n$ as the
  phenomenological FRDM.
  For each isobaric chain with mass $A$, the distance between the nuclide $(Z,A)$ and the nuclide
  $(Z_0,A)$ in the $\beta$-stability line~\cite{MS71} is defined by $\varepsilon=Z_0-Z$ with
  \begin{eqnarray}
   \label{eq:stable}
   Z_0=\frac{A}{1.98+0.0155A^{2/3}} \; ,
  \end{eqnarray}
  i.e., $\varepsilon=0$ stands for the most stable nuclei and
  $\varepsilon>0$ the neutron-rich nuclei. The $rms$ deviation
  $\sigma_{rms}$ of $S_n$ as a function of $\varepsilon=Z_0-Z$ for
  different mass models is shown in Fig.~\ref{fig:sn_sigma}. It is
  remarkable that almost the same order of prediction power of $S_n$
  from the neutron-deficient side to the neutron-rich side is
  achieved for all the models, even though models like FRDM and
  HFB-13 have about 10 more free parameters than the RMF model and
  were optimized for all the known masses.
  While the macroscopic-microscopic mass model FRDM shows the best
  agreement with experimental values in the neutron deficient mass
  region, it gets progressively worse when moving away from the
  stability line towards the neutron-rich side.

\begin{figure}
  \includegraphics[scale=0.8]{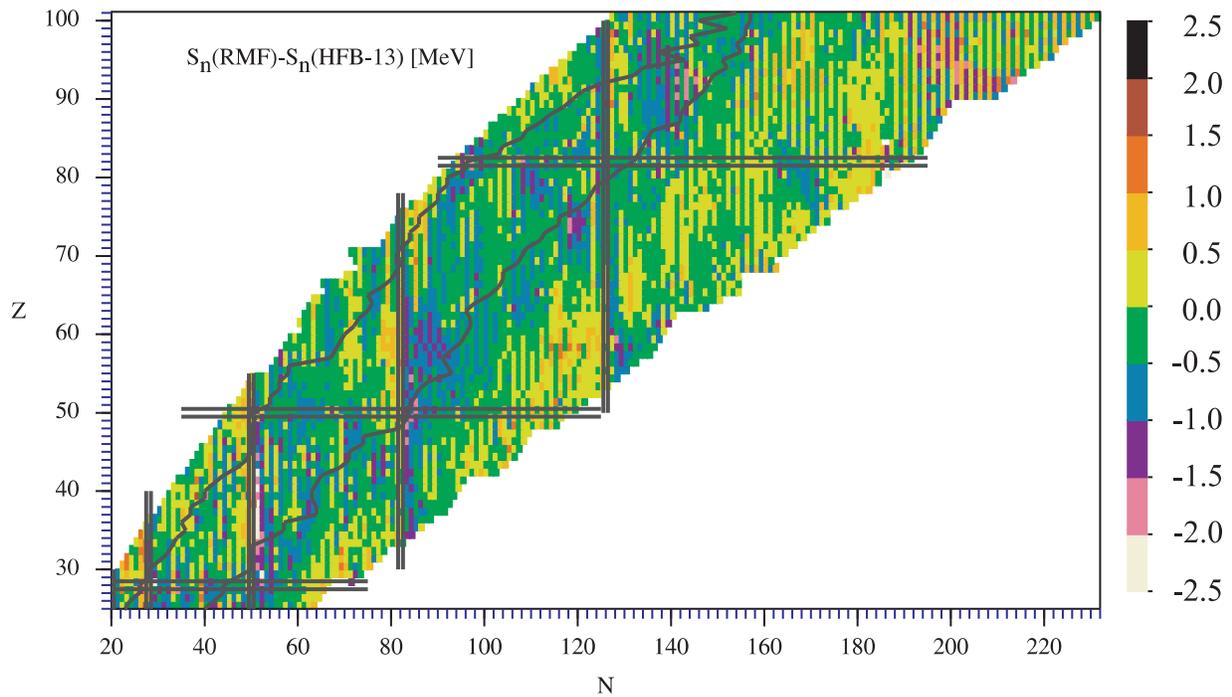}
  \caption{\label{fig:snd} \baselineskip=13pt
  (Color online) The differences between one neutron separation energies $S_n$ predicted in the RMF model and those in the
  HFB-13 model. The magic proton and neutron numbers are indicated by pairs of parallel lines, and
  also the present border of the data with known masses are shown by solid lines.}
\end{figure}

  All theoretical models addressing nuclei far away from the $\beta$-stability line involve a
  dramatic extrapolation to unknown nuclei. Thus it is interesting to examine what is the difference
  of $S_n$ predicted in the different models when going towards the neutron-rich side.
  In Fig.~\ref{fig:snd}, the differences between $S_n$ in the RMF model and those in
  the HFB-13 mass models are shown as an example. In general, most
  of the discrepancies between the two microscopic models range from -1
  MeV to 1 MeV across the entire nuclear chart. Furthermore, the
  $S_n$ value is consistent with that in the HFB-13 model in the
  range of -0.5 MeV to 0.5 MeV when going to the unknown region of the nuclei chart.
  These differences indicates that the extrapolation can be quite different depending on
  the underlying physics of the model. Similar conclusion can be drawn also
  for the comparison of the RMF and FRDM models.
  Around the $N=82$ shell, the RMF model predicts a systematic lower $S_n$.
  These different $S_n$ predictions towards the neutron drip
  line affect $r$-process calculations and thus the corresponding determined
  astrophysical conditions.

\begin{figure}
  \includegraphics[scale=0.8]{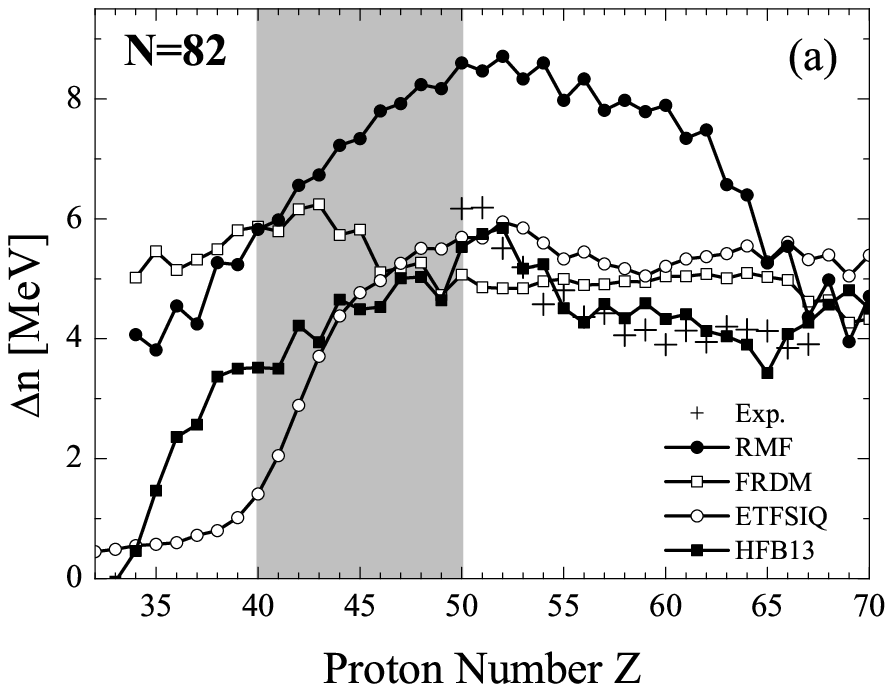}%
  \includegraphics[scale=0.8]{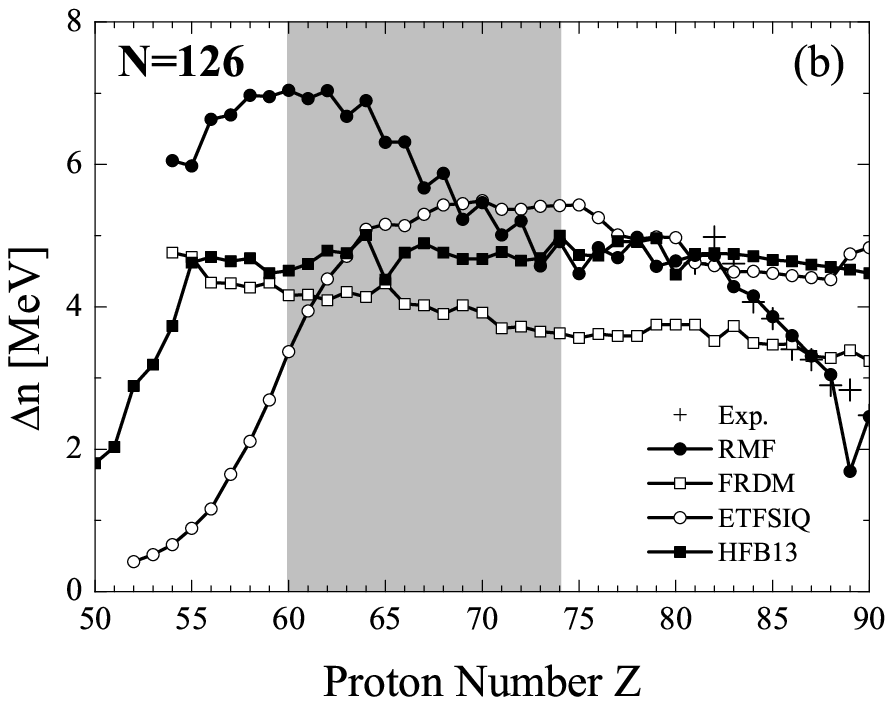}
  \caption{\label{fig:gap} \baselineskip=13pt The neutron shell gaps
  $\Delta_n(Z, A)= S_{2n}(Z, A)-S_{2n}(Z, A+2)$
  for $N = 82$ and 126 in the RMF approach compared with those in the FRDM, ETFSI-Q and HFB-13 models
  together with the data available.
  The nuclei in the shadowed areas are involved in the $r$-process paths based on our calculations.}
\end{figure}

  The evolution of the nuclear structure around the shell closures $N = 82$ and 126 is critical in
  understanding the $r$-process abundance distribution around the $A = 130$ and $A = 195$ abundance
  peaks. At the shell closures, the one-neutron separation energy drops, and thus the corresponding
  nucleus in the $r$-process path cannot absorb another neutron without photo-disintegration. Therefore
  it has to ``wait'' for the $\beta$-decay to proceed, and the path moves closer to the valley of
  stability where the half-lives are longer. These isotopes with long half-lives serve as
  bottlenecks of the process where abundances accumulate and the abundance peaks are formed. In
  Fig.~\ref{fig:sn82}, the predicated average one-neutron separation energy $S_{2n}/2$ around $N =
  82$ in the RMF model is displayed as a function of mass number for isotopes from Kr to Ba together
  with the experimental values and the predictions of the FRDM, ETFSI-Q and HFB-13 models.
  The dominating isotopes in the $r$-process path (defined as waiting points and discussed in
  Sec.~\ref{sec:calculation}) are also
  indicated for the corresponding mass model. Similarly, Fig.~\ref{fig:sn126} presents $S_{2n}/2$
  distribution around $N = 126$ shell for isotopes from Ce to Pt. In general, the RMF model
  reproduces well the experimental data, and predicts a much subtle variance relative to other mass
  models. The neutron shell gaps, defined as $\Delta_n(Z,A) = S_{2n}(Z,A)-S_{2n}(Z,A+2)$, can be more
  clearly seen from Fig.~\ref{fig:gap}, which shows the shell gaps for $N = 82$ and 126 in the RMF
  approach in comparison with the data available and those in the FRDM, ETFSI-Q and HFB-13 models.
  The nuclei in the shadowed area are in the $r$-process path. At the $N = 82$ shell, all the mass
  models except the FRDM model show a strong quenching effect (i.e., the shell gap drops) towards the
  neutron-rich side. The RMF shell gap is overestimated compared with the data available and it is
  around 2 MeV larger than other models for $45 \leq Z \leq 60$.
  Regardless of this, the RMF model succeeds in predicting the
  enhanced double-magic effect at $Z=50$ together with the HFB-13 model.
  For $N = 126$, there is no sign of shell quenching observed in the $r$-process region for all the models.
  A unique feature of the RMF model is that it fully coincides with the available data
  and it is also the only model to reproduce the
  enhanced double-magic effect at $Z=82$. In comparison, the other models
  fail to reproduce the trend of the known $N=126$ shell. Towards
  the neutron drip line, the RMF prediction tends to enhance the
  shell until the maximum is reached around $Z=60$ while the other models have a roughly
  constant shell gap.

\section{Site-independent $r$-process approach}
  \label{sec:$r$-process}
  Since the $r$-process is responsible for the synthesis of half the heavier nuclei beyond the iron group~\cite{Cowan-PR91, Qian-PPNP03},
  the basis of the nucleosynthesis mechanism have been
  extensively studied. Nevertheless, the location where it occurs has not been unambiguously
  identified. Current location candidates include the neutrino-driven wind off a proton-neutron
  star in core collapse supernovae \cite{WWH92,TWJ94,Wan01,Tho01}, neutron star mergers
  \cite{FRT99,ROS99,GDJ05}, jets in core collapse supernovae \cite{Cam01}, shocked surface layers
  of O-Ne cores \cite{Ning07}, and gamma ray bursts \cite{SuM05}. Because the specific
  astrophysical conditions among the different scenarios may change, solar $r$-process
  abundances~\cite{Cowan-06} have been used in the past to constrain the astrophysical conditions
  using a site-independent approach~\cite{Kartz-APJ93,kratz06}. In this approach seed-nuclei
  (usually the iron group) are irradiated by neutron sources of high and continuous neutron densities
  $n_n$ ranging from $10^{20}$ to $10^{28}$ cm$^{-3}$ over a timescale $\tau$ in a high temperature
  environment ($T \sim$ 1GK). This superposition of $r$-process components ($n_n$,$\tau$) is needed to
  reproduce the overall shapes and positions of the solar $r$-process
  abundances~\cite{Kartz-APJ93,pfeiffer-ZPA97,CPK-APJ99} and it is equivalent to the exponential
  neutron exposures in the $s$-process~\cite{KBW-RPP89}. The configuration of many $r$-process
  components seems to be also a reasonable approximation to the real $r$-process event. For instance, one can
  think of it as the ``onion" structure of neutron sources with different densities, where the
  seed-nuclei capture neutrons while moving through different zones with different thicknesses. In
  this paper we explore for the first time the application of the new RMF mass model to an $r$-process
  calculation and at the same time investigate the effect of nuclear physics uncertainty in the $r$-process.

  Due to the high neutron densities, neutron captures are much faster than the competing
  $\beta$-decays and an (n,$\gamma$)$\Leftrightarrow$($\gamma$,n) equilibrium is nicely established
  for every element. The abundance ratio of two isotopes in the timescale $\tau$ can be expressed simply as
  \begin{eqnarray}
    \frac{Y(Z,A+1)}{Y(Z,A)} &=& n_n \left(\frac{h^2}{2\pi m_\mu\kappa T}\right)^{3/2}
    \frac{G(Z,A+1)}{2G(Z,A)}\left(\frac{A+1}{A}\right)^{3/2} \cr
          & & \exp\left[\frac{S_n(Z,A+1)}{\kappa T}\right] \; ,
    \label{eq:saha}
  \end{eqnarray}
  where $Y(Z,A)$ denotes the abundance of the nuclide $(Z,A)$, $S_n$
  is the one-neutron separation energy, $G(Z,A)$ is the partition
  function of nuclide $(Z,A)$, and $h$, $\kappa$ and $m_\mu$ are
  the Planck constant, Boltzmann constant and atomic mass unit,
  respectively. Neglecting the difference in the ratios of the
  partition functions and the atomic mass, one can easily see that
  the isotopic abundance distribution $P(Z,A)$ and the abundance
  maxima in each isotopic chain are determined by $n_n$, $T$ and
  $S_n$. Approximating $Y(Z,A+1)/Y(Z,A) \simeq 1$ at the highest
  isotopic abundance for each element, and all other quantities
  being constant, the average neutron-separation energy $\bar{S}_n$,
  calculated by
  \begin{eqnarray}
    \bar{S}_n &\approx& \kappa T \log\left[\frac{2}{n_n}\left(\frac{2\pi m_\mu\kappa
    T}{h^2}\right)^{3/2}\right] \cr
     &=& T_9 \left\{2.79 + 0.198\left[\log
     \left(\frac{10^{20}}{n_n}\right)+\frac{3}{2}\log T_9\right]\right\} \;,
     \label{eq:sn}
  \end{eqnarray}
  is the same for all the nuclides with the highest abundance in each isotopic chain.
  $T_9$ denotes the temperature in $10^9$ K. Higher temperature or
  lower neutron density will drive the $r$-process path towards the
  valley of stability. Due to the pairing correlation the most
  abundant isotope always has an even neutron number $N$.

  If fission is neglected, the abundance flow from one isotopic chain to the next is governed by $\beta$-decays
  and can be expressed by a set of differential equations:
  \begin{eqnarray}
     \frac{dY(Z,A)}{dt} &=&  Y(Z-1)\sum_AP(Z-1,A)\lambda_\beta^{Z-1,A}  \cr
      & & - Y(Z)\sum_A P(Z,A)\lambda_\beta^{Z,A} \;,
     \label{eq:flow}
  \end{eqnarray}
  where $\lambda_\beta^{Z,A}$ is the total decay rate of the nuclide $(Z,A)$ via the $\beta$-decay and
  the delayed neutron emission, and
  $Y(Z)=\sum_AY(Z,A) = \sum_A P(Z,A)Y(Z)$ is the total abundance in each isotopic chain.
  Using Eqs.~(\ref{eq:saha}) and (\ref{eq:flow}), the abundance for each isotope can be calculated.
  After the neutrons freeze out, all the isotopes will proceed to the corresponding stable isotopes via $\beta$-decays.

\section{Calculations}
    \label{sec:calculation}
  In the present calculation, unknown one-neutron separation energies $S_n$ were calculated from the
  RMF approach~\cite{RMFBCS} and $\beta$-decay properties were taken from Ref.~\cite{Moller03}.
  Available experimental data \cite{AW03,NNDC} was used when available.
  Similar to the method used in Refs.~\cite{pfeiffer-ZPA97,CPK-APJ99,kratz06},
  we applied sixteen components with neutron densities in the range of
  $10^{20}$-$3\times10^{27}$ cm$^{-3}$ in our calculation. We chose a temperature $T=1.5$ GK. We
  assumed that for this temperature the irradiation time $\tau$ and the corresponding weight $w$
  follow the exponential dependent of neutron density $n_n$, i.e.,
  \begin{equation}
    w(n_n)=n_n^a , \; \tau(n_n)=b \times n_n^c   \;,
    \label{eq:power}
  \end{equation}
  where $a, b, c$ are parameters to be fixed. These parameters can be obtained from a least-square
  fit to the solar $r$-process abundances. We further assume that the longest neutron irradiation
  time has to be longer than 0.5 s but shorter than 20 s. The exponential relations in
  Eq.~(\ref{eq:power}) have been observed when fitting the three $r$-process peaks~\cite{Chen-PLB95}
  and used for stellar and chronometers studies~\cite{pfeiffer-ZPA97,CPK-APJ99,Schatz-APJ02}.

\begin{figure}
  \includegraphics[scale=1]{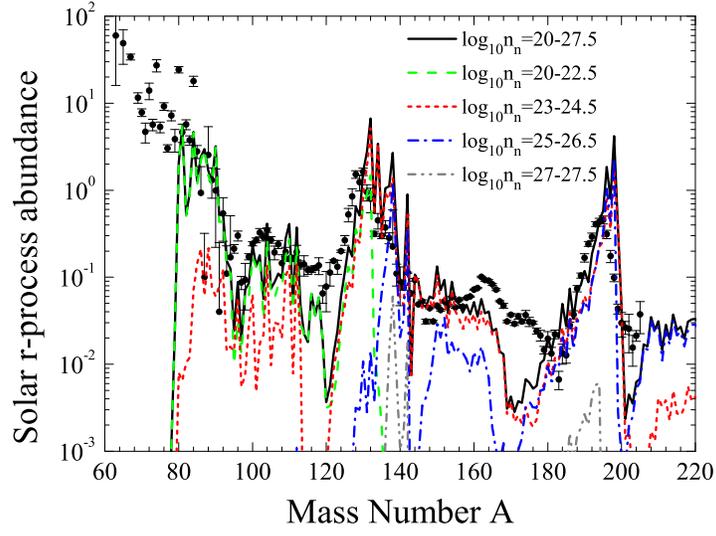}
  \caption{\label{fig:nn_component}\baselineskip=13pt
   (Color online) The effect of various weighted $r$-process components
   on the resulting fit after $\beta$-decays in the best
   superposition using the RMF masses. The calculated total isotopic
   abundances (in the logarithm scale)) is normalized to $A = 130$.
   }
\end{figure}

  It was found that $r$-process components with $\tau(n_n) = 0.454 n_n^{0.040}$ s and $w(n_n) = 2.1
  n_n^{0.02}$ best reproduce the solar $r$-process abundance. Fig.~\ref{fig:nn_component} shows the
  contribution of the four-group weighted $r$-process components after $\beta$-decays to the resulting
  best fit. The black solid curve with isotopic abundances normalized to $A = 130$, corresponds to
  the fit using all the sixteen components. The green, red, blue and grey dashed curves are the sum
  of the abundances calculated with log$(n_n)$ ranging from 20 to 22.5, 23 to 24,5, 25 to
  26.5 and 27 to 27.5, respectively. The first six components with log$(n_n)$ between 20 and 22.5
  seem to account for the $A=80$ abundance peak. The four components with log$(n_n)$ between 23 and
  24.5 are responsible for the overall structure of the $r$-abundance curve beyond $A=120$ and the
  remaining components only improve the description of the theoretical calculation for $A>150$. In
  general, the fit is found to reproduce well the solar $r$-process abundances and also the
  position of the abundance peaks.

  The $r$-process runs relatively close to the $\beta$-stability line around the shell closure, thus
  the experimentally known mass values around $N=82$ shell significantly influence the
  abundance distribution after the second abundance peak. Taking our best simulation using the RMF
  masses as an example, the ratio between the abundance at $A=130$ and the abundance at $A=195$ is
  2.8 when taking the experimental data into account, and it increases to 24.6 if the
  experimental data are not used. The same ratio decreases from 4.0 to 1.3 for the best
  simulation using the FRDM masses (to be discussed below).
  In order to minimize the contribution from the theoretical
  uncertainty of known masses, experimental information was included
  in the calculations.

\begin{figure}
  \includegraphics[scale=1]{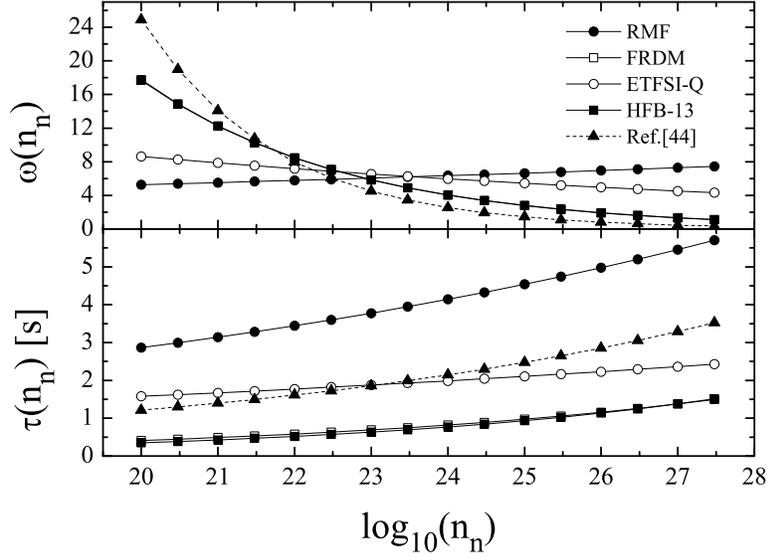}
  \caption{\label{fig:RMF_condition} \baselineskip=13pt
  The best configuration of sixteen $r$-process components which reproduce the solar
  system $r$-process abundances with different mass inputs at the temperature $T=1.5$ GK.
  The neutron density $n_n$ is in the unit of cm$^{-3}$.
  The weighting factor $\omega(n_n)$ and the neutron irradiation time $\tau(n_n)$ are shown in the upper and
  lower panels as a function of neutron density $n_n$.
  In the upper panel, the weighting factors for the FRDM are completely overlaid by those for the HFB-13 models.
  The fit from Ref.~\cite{CPK-APJ99} is also plotted
  for comparison. The total weighting factor has been normalized to 100.}
\end{figure}

  In order to investigate the impact of theoretical uncertainty of unknown masses in
  an $r$-process calculation, we also performed the same procedure using instead the FRDM,
  ETFSI-Q and HFB-13 mass predictions while keeping the same
  $\beta$-decay properties.
  The astrophysical conditions determined by using various mass inputs are shown in
  Fig.~\ref{fig:RMF_condition}.
  The obtained superpositions of sixteen $r$-process components
  for all the mass tables are collected in table~\ref{tab:fit}.
  Similar to the astrophysical conditions obtained from ETFSI-Q simulations, the astrophysical
  condition using the RMF mass input requires a relatively constant weighting factor for different
  neutron densities. However, the FRDM and HFB-13 cases
  favor a large weighting factor for the low neutron density. As for the neutron irradiation time,
  the best RMF fit requires component durations of as long as 6 s while
  the FRDM and HFB-13 simulations only requires up to
  1.5 s. The ETFSI-Q component durations are somewhat in between.
  Moreover, it may be worth mentioning that the simulations using the FRDM and HFB-13
  masses demand almost identical astrophysical conditions.
  Using a lower temperature $T =1.35$ GK, a similar
  calculation based on the ETFSI-Q mass model is carried out in Ref.~\cite{CPK-APJ99}.
  As shown in Fig.~\ref{fig:RMF_condition}, their
  obtained neutron irradiation times are in good agreement with our calculation
  using the ETFSI-Q masses and FRDM half-live
  inputs, but the weighting factors differ.
  The superposition obtained in Ref.~\cite{CPK-APJ99} demands a more sharp evolution of
  the weighting factor as a function of $n_n$.
  Since a lower temperature of $T =1.35$ GK in our calculations only weakly impacts the
  condition obtained, the difference should be due to the different $\beta$-decay properties
  used in that work.
  Based on a full dynamical network calculation, faster time-scales of the order of
  hundreds of milliseconds, are found in Ref.~\cite{farouqi-NPA05} for an $r$-process
  in the neutron-wind scenario of core-collapse type II
  supernovae. However, this different time scale can be at least partially
  attributed to different seed nuclides. In their calculation, the
  $r$-process stars from a seed distribution containing neutron-rich
  nuclei with mass numbers between 80 and 100, while ours starts from
  $^{56}$Fe.

\begin{figure}
  \includegraphics[scale=1.5]{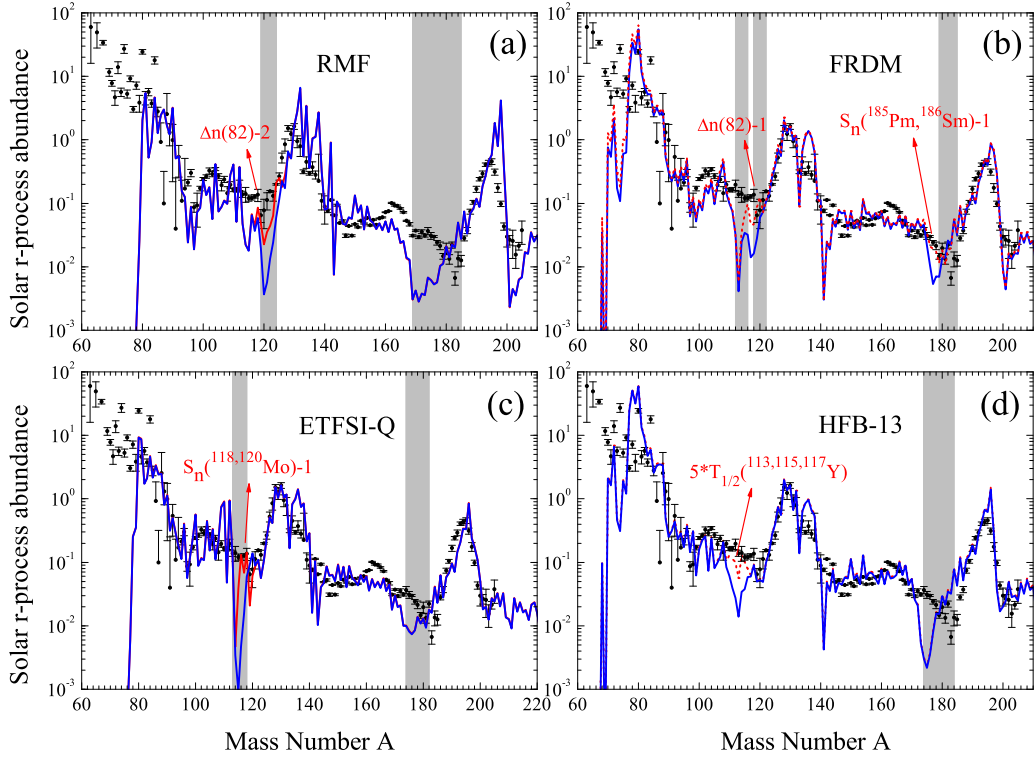}
  \caption{\label{fig:best} \baselineskip=13pt
  (Color online) Our best fits to the solar $r$-process abundances (in the logarithm scale) under different
  nuclear mass inputs. The $\beta$-decay properties are taken from the FRDM model~\cite{Moller03}.
  The best fits are displayed as blue-solid lines.
  In the sub-figure (a), the red-dashed curve is the same as the blue-solid curve but with
  a shell closure at $N=82$ 2 MeV smaller.
  In the sub-figure (b), the red-dashed curve is the same as the blue-solid curve but with
  a shell closure at $N=126$ 1 MeV smaller and separation energies of $^{185}$Pm and $^{186}$Sm 1 MeV smaller.
  In the sub-figure (c), the red-dashed curve is the same as the blue-solid curve but
  with separation energies of $^{185}$Pm and $^{118,120}$Sn 1 MeV smaller.
  In the sub-figure (d), the red-dashed curve is the same as the blue-solid curve but with
  half-live of isotopes $^{113,115,117}$Y five times larger .
  The shadowed areas correspond to the range where the abundances of these isotopes
  are largely underestimated before neutrons freeze out. }
\end{figure}

  Calculated solar $r$-process abundances after $\beta$-decays using
  different mass models are displayed in Fig.~\ref{fig:best}. Shadowed areas show the regions
  with underproduced abundances
  before the neutrons freeze-out. After $\beta$-decays to the stability line,
  these gaps are too large to be completely filled in by $\beta$-delayed neutron emissions.
  It should be pointed out that the solar system $r$-process abundances are defined as the abundances not
  produced in the $s$-process and $p$-process that still have to be created elsewhere to explain the
  solar system abundances. Although it is thought that the $r$-process is responsible for the majority
  of those isotopic abundances with $Z \geq$ 56, its contribution to the lighter elements is still
  debatable~\cite{Travaglio-APJ04,Montes07}. It is possible that some of the discrepancies in the
  reproduction of the low mass abundances may be due to an additional nucleosynthesis component
  creating some of those abundances. However, since astrophysical conditions and nuclear properties
  both affect the resulting r-process abundances, one need to determine or at least understand
  the uncertainty in the nuclear physics properties in the future works to disentangle both effects~\cite{SM08CPL}.
  In this paper we only discuss possible nuclear physics reasons for
  such underabundances.

\begin{figure}
  \includegraphics[scale=1.6]{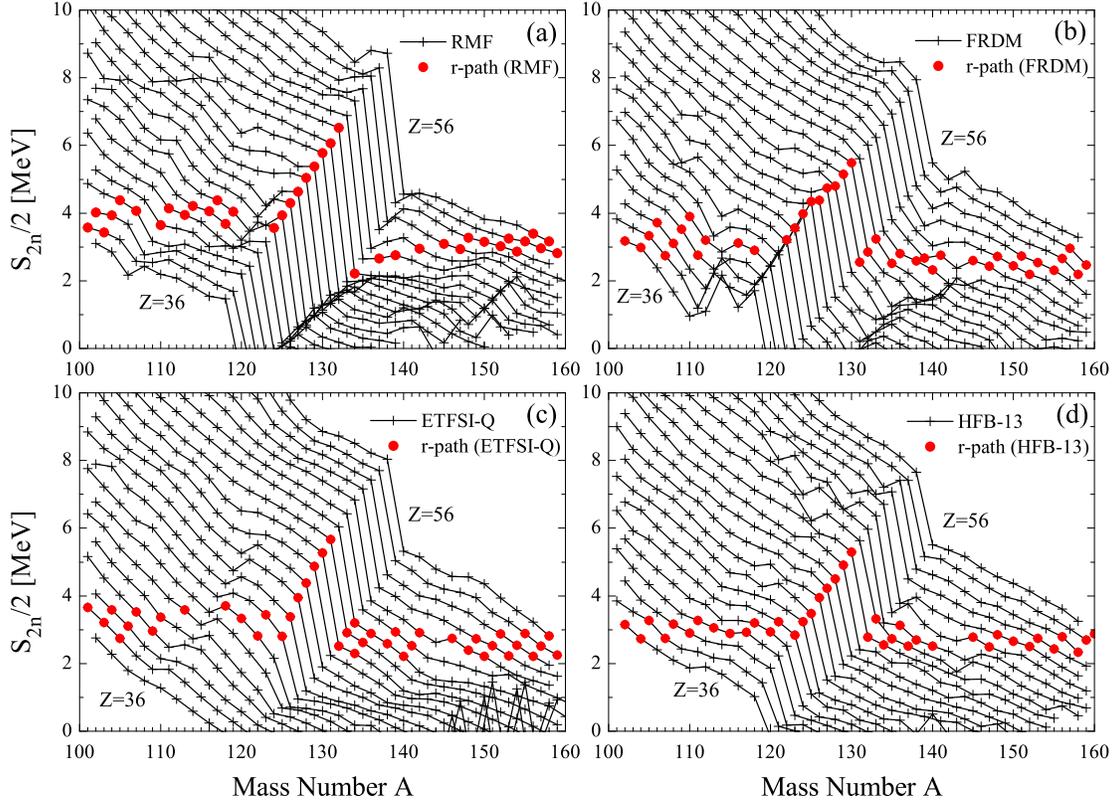}
  \caption{\label{fig:sn82} \baselineskip=13pt
  (Color online) The average one-neutron separation energies around the $N = 82$ shell in the RMF model,
  in comparison with those in the FRDM, ETFSI-Q and HFB-13 models as a function of mass
  number $A$. For simplicity only nuclei with even $N$ are
  plotted. The corresponding $r$-process paths calculated using different mass inputs are also
  indicated by dots, and labeled here are those isotopes with more than 10\% population of each isotopic chain.}
 \end{figure}

\begin{figure}
  \includegraphics[scale=1.6]{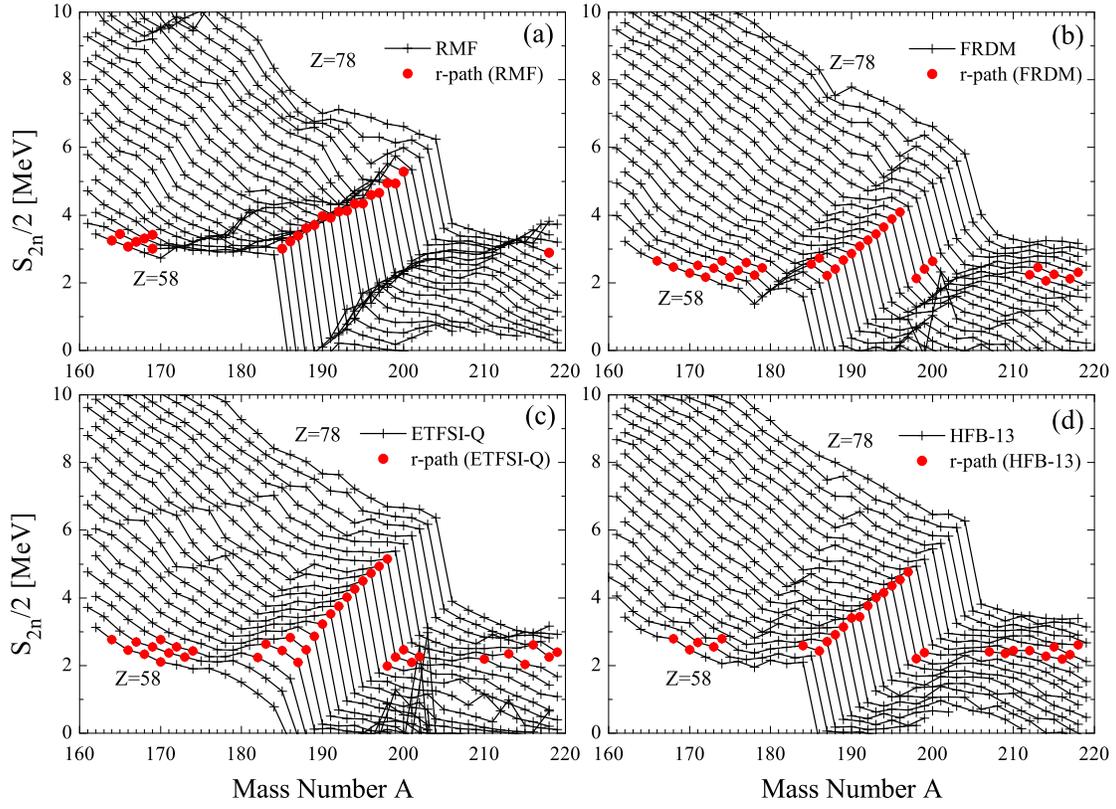}
  \caption{\label{fig:sn126} \baselineskip=13pt
  (Color online) Same to Fig.~\ref{fig:sn82} but around the $N = 126$ shell.}
\end{figure}

$R$-process abundances calculated with all nuclear mass models
result in abundance underproduction at $A \sim$ 120 and $A \sim$
170. Traditionally, the underestimation of the isotopic abundances
before $A \sim $ 130 peak has been attributed to the overestimated
strength of the $N$ = 82 shell closure~\cite{Chen-PLB95,
pfeiffer-ZPA97, CPK-APJ99} in the theoretical nuclear physics model
even though the experimental evidence is still
debated~\cite{dillmann03,walters04,rising07}. Since it is not
possible to do a complete study of the shell-quenching effect at the
single particle level which should affect more nuclei than the one
with $N=82$, we only study the effect of a reduced shell closure by
artificially decreasing the shell gap energies at $N=82$ in the RMF
and FRDM models by 2 MeV and 1 MeV, respectively. In such a way, the
shell gaps interested for the $r$-process would roughly have the
same values as those in the quenched models ETFSI-Q and HFB-13.
Eventually, a better agreement with the observation at $A\sim 120$
is obtained as shown in Fig.~\ref{fig:best}(a-b). This can be easily
understood as follows. A reduction of shell gap leads to a nuclear
matter repopulation in the isotopic chain according to
Eq.~\ref{eq:saha}. $R$-process waiting points located at $N=82$ move
closer to the valley of stability and thus some of the
underabundance can be filled. Furthermore, based on
Fig.~\ref{fig:sn82} one could expect that the quenched shell gap at
$N=82$ would not affect the abundance around $A\sim 115$. It is
interesting to note that the $r$-process simulation using the
shell-quenched ETFSI-Q model in Fig.~\ref{fig:best}(c) show a good
agreement with solar abundances pattern at $A\sim 120$ together with
a large underproduction at a lower mass number $A\sim 115$.

The abundance trough at $A\sim 115$ for the FRDM, RMF and ETFSI-Q
models can be related to the additional bump of $S_n$ at $A=110-120$
in Fig.~\ref{fig:sn82}, and thus the associated nuclear shape
transition. In the case of the ETFSI-Q model, nuclear shape changes
for prolate to oblate and then to spherical nuclei with $N$ = 82.
This transition leads to a deviation from the approximate
relationship between neutron separation energies and mass number for
each isotope, and can be clearly recognized in
Fig.~\ref{fig:sn82}(c) by the sudden increase of the separation
energies. In order to see the sensitivity of the $r$-process
calculation to the effect of the nuclear shape transition, we
lowered the separation energies of $^{118,120}$Mo by 1 MeV in the
ETFSI-Q model, but kept the other nuclear physics input unchanged.
Those isotopes are in the $r$-process path and show a bump in the
one-neutron separation energies as a function of mass number. As
shown in Fig.~\ref{fig:best}(c) the $A \sim 115$ trough is largely
filled in. A similar analysis shows the same conclusion for the FRDM
and RMF mass models. Although we have mainly focused in the
underabundance below $N=82$, similar conclusion can be drawn for the
trough around $A \sim 170$. As an example, Fig.~\ref{fig:best}(c)
shows that the trough is almost completely filled in by lowering the
separation energies of $^{185}$Pm and $^{186}$Sm by 1 MeV. This
suggests that the potentially wrongly assigned location of the shape
transition before the neutron magic number in the theoretical
predictions can lead to the troughs before the abundance peaks.

Of all the mass models,  the HFB-13 is the only one that shows a
smooth one-neutron separation energy change from Sr to Ru. As a result,
the $r$-process waiting points are continuous and there is
not apparent gap in the $r$-process path (see
Fig.~\ref{fig:sn82}(d)). Only modifications in the nuclear masses
would not result in the filling of the the abundance gap at $A\sim
115$. Such underproduction may be traced back to the $\beta$-decay properties.
By increasing the $\beta$-decay half-lives
of the critical nuclei $^{113,115,117}$Y by five times, we found
that the trough before the $A \sim 130$ peak in the HFB-13 case can
be nearly filled in as shown in Fig.~\ref{fig:best}(d).

\section{\label{sec:conclusion} Summary}

  We have applied the most recent comprehensive mass
  models, the non-relativistic microscopic-rooted HFB-13
  and the relativistic RMF in $r$-process calculations. For the sake of comparison, we also included the widely used
  macro-microscopic models FRDM and ETFSI-Q. Of these models, the HFB-13 and RMF models are
  used for the first time in such calculations. Based on a simple $r$-process model, it is found that all mass
  models reproduce the main features of the solar $r$-process pattern and the position of the abundance peaks.
  Since $r$-process simulations have to rely on predicted nuclear physics properties of unknown
  regions in the nuclear chart, we have compared the predictions of different mass models. We have also made a systematic
  study of the influence due to the mass model uncertainty in the application of the $r$-process
  and thus in the required astrophysical conditions. This nuclear
  physical uncertainty is very important for the complete understanding of the $r$-process
  since the results of more modern full dynamic $r$-process calculations depend on the nuclear mass
  input used. It is found that the deduced astrophysical conditions like the neutron irradiation
  time of the $r$-process can be significantly different depending on the mass model
  used. Among the different models, the simulation using the RMF masses requires a
  longer time scale (up to a factor of 4) than those using FRDM and HFB-13
  models. Furthermore, it is found that the optimal astrophysical conditions
  obtained using the ETFSI-Q and RMF mass models require a
  relatively constant weighting factor for neutron densities in the range
  $10^{22}$ to $10^{28}$ cm$^{-3}$, while
  the FRDM and HFB-13 simulations favor a large weighting factor at low densities.
  In addition, we have explored the possible deficiencies
  in different mass models, and found that the observed abundance
  underproduction before the abundance peaks in all the models can be a
  combined and complex effect of both shell structure and shape
  transition. An exception is the underproduction at $A\sim
  115$ in the HFB-13 model which can be attributed to incorrect
  $\beta$-decay rates. Future experiments are needed to determine the strength of
the shell closure towards the neutron drip line as well as the precise locations of the shape transition toward the shell-closures.

\begin{acknowledgments}
 We thank C. Scheidenberger for his interest of this work, G. Martinez Pinedo and B. Pfeiffer for valuable
 discussions, J.M.~Pearson for providing the ETFSI-Q and HFB-13 mass tables and P.~M\"{o}ller for the $\beta$-decay data.
 This work is partly supported by Major State Basic Research
 Developing Program 2007CB815000 as well as the National Natural Science Foundation of China under
 Grant No. 10435010, 10775004 and 10221003.
\end{acknowledgments}

\clearpage

\clearpage

\clearpage

\clearpage

\clearpage

\clearpage

 \begin{table}[h!]
 \centering
 \tabcolsep=6pt
 \caption{\label{tab:fit} Our best
fits to the solar $r$-process abundances for different sets of
nuclear mass models. The first column is the mass model employed.
The last two columns are the weight $\omega$ and the relevant
neutron irradiation time $\tau$ (in unit of second), respectively. }
 \begin{tabular}{ccc}
 \hline
Mass model & {\mbox{$\omega$}} &  {\mbox{$\tau$}}~[s]\\\hline
RMF        &  2.1$\times n_n^{0.020}$     & 0.454$\times n_n^{0.040}$  \\
FRDM       &  3.0E4$\times n_n^{-0.161}$  &  0.013$\times n_n^{0.075}$ \\
ETFSI-Q    &  54.4$\times n_n^{-0.040}$   & 0.499$\times n_n^{0.025}$  \\
HFB13      &  2.8E4$\times n_n^{-0.160}$  &  0.007$\times n_n^{0.085}$ \\
 \hline
  \end{tabular}
\end{table}

\clearpage

\end{document}